# Design and Performance Analysis of an Ultra-Low Power Integrate-and-Fire Neuron Circuit Using Nanoscale Side-contacted Field Effect Diode Technology

SEYEDMOHAMADJAVAD MOTAMAN, SARAH S. SHARIF, AND YASER M. BANAD

School of Electrical and Computer Engineering, University of Oklahoma, Norman, OK 70803, USA

CORRESPONDING AUTHOR: YASER M. BANAD (e-mail: bana@ou.edu).

**ABSTRACT** Enhancing power efficiency and performance in neuromorphic computing systems is critical for next-generation artificial intelligence applications. We propose the Nanoscale Side-contacted Field Effect Diode (S-FED), a novel solution that significantly lowers power usage and improves circuit speed, facilitating efficient neuron circuit design. Our innovative integrate-and-fire (IF) neuron model demonstrates exceptional performance metrics: 44 nW power consumption (85% lower than current designs), 0.964 fJ energy per spike (36% improvement over state-of-the-art), and 20 MHz spiking frequency. The architecture exhibits robust stability across process-voltage-temperature (PVT) variations, maintaining consistent performance with less than 7% spike amplitude variation for channel lengths from 7.5nm to 15nm, supply voltages from 0.8V to 1.2V, and temperatures from -40°C to 120°C. The model features tunable thresholds from 0.8V to 1.4V and reliable operation across input spike pulse widths from 0.5 ns to 2 ns. This significant advancement in neuromorphic hardware paves the way for more efficient brain-inspired computing systems.

## I. INTRODUCTION

Conventional processors, particularly those following the von Neumann architecture, consist of distinct components such as the Arithmetic Logic Unit (ALU), Control Unit (CU), and registers, each consuming substantial energy. These components individually contribute to the overall power consumption of the computing system, increasing energy inefficiencies. A core characteristic of the von Neumann model is the separation of processor and memory units, requiring continuous data exchanges for computational tasks. This structural design creates significant inefficiencies due to frequent data transfers, known as the von Neumann bottleneck [1–4].

To address the high energy demands of traditional computing systems, researchers have been exploring alternative paradigms, with neuromorphic computing emerging as a leading approach. Inspired by the human brain's operational mechanisms, neuromorphic systems aim to replicate the brain's efficiency in executing complex tasks—such as pattern recognition, reasoning, and decision-making—using approximately 20W of power [5]. In contrast to the von Neumann architecture, neuromorphic systems leverage highly efficient, parallel processing capabilities to handle sophisticated computations like clustering, classification, and pattern recognition with significantly reduced energy consumption [6,7].

The brain's computational ability lies in its network of billions of interconnected neurons and synapses, which form the basis of cognition and processing [8]. This biological model inspires silicon-based spiking neural networks (SNNs), where neurons and synapses are mimicked using advanced electronic components. Emerging memory technologies, such as Phase Change Memory (PCM) and Resistive Random Access Memory (RRAM), are particularly promising for emulating synaptic weights while reducing power consumption compared to CMOS-based designs [6,7,9–11]. Similarly, the development of efficient neuron circuits has been a focus, as these are crucial for minimizing the overall power and energy requirements of neuromorphic systems while enhancing hardware performance for complex computations.

Neuron circuit designs have evolved significantly over the past two decades improving the limitations of earlier models by enhancing efficiency, reducing power consumption, and improving operational speed. For example, in 2003, Indiveri introduced a CMOS integrate-and-fire (IF) neuron circuit that consumed only 1.5 μW of power. This marked an important milestone in energy-efficient neuromorphic systems, but its limited spiking frequency hampered overall performance [12]. In recent years, numerous advancements have been made in neuron circuit design, each targeting specific aspects to more closely replicate the efficiency and functionality of biological neurons.

In 2016, Tuma et al. developed a neuron circuit using nanoscale phase-change devices to implement membrane potentials, achieving high area efficiency. However, this design consumed 5 pJ per spike, making it less efficient than other approaches [13].

In 2017, Dutta et al. proposed a partially depleted SOI-MOSFET design that utilized the floating body effect to mimic leaky integrate-and-fire (LIF) neuron behavior. While this architecture operated at frequencies in the MHz range, its energy consumption of 35 pJ per spike limited its efficiency [14]. By 2018, Kyu-Bong Choi et al. introduced an innovative IF neuron circuit based on a Split-Gate Positive Feedback Device combined with CMOS technology. This design significantly improved energy consumption, achieving a remarkable 0.25 pJ per spike, which was 100 times more efficient than the best designs at the time [15].

In 2019, Chatterjee et al. introduced bulk FinFET-based neuron circuits, demonstrating an energy efficiency of 6.3 fJ per spike and achieving a spiking frequency of 2 MHz. This innovation set a new benchmark for low-power neuromorphic systems [16]. The field saw further advancements in 2020, with Chavan et al. leveraging BTBT-induced hole storage in SOI-MOSFETs to achieve an energy consumption of 3.22 fJ per spike. Additionally, Woo et al. implemented an IF neuron model using Feedback Field-Effect Transistors (FBFETs), which consumed 2.9 fJ per spike at a spiking frequency of 20 kHz [17].

Recent years have seen even more groundbreaking developments. In 2023, Dongre et al. introduced an RRAM-based neuron circuit that incorporated a control unit to optimize power consumption. This architecture achieved an energy efficiency of 1.5 fJ per spike and operated at frequencies ranging from 277 kHz to 3 MHz [18]. In 2024, Zuber Rasool et al. proposed a novel LIF neuron circuit utilizing a Trench Gate Vertical Feedback Field-Effect Transistor (TG-V-FBFET) with a positive feedback mechanism. This circuit reached a spiking frequency of 860 MHz while consuming 130 fJ per spike [19].

Touchaee et al. examined the potential of side-contacted field-effect diodes (SFEDs), which exhibited a high on/off current ratio and low parasitic capacitance. These attributes substantially reduced the power-delay product, enabling efficient and high-speed operation with minimal energy consumption [20-21]. In our earlier research [27], we briefly introduced the architecture of an S-FED-based IF neuron circuit. Building on this research, this paper investigates the nanoscale S-FED as a cornerstone for neuromorphic computing by presenting several significant contributions to the field. First, we have developed a novel nanoscale S-FED-based integrate-and-fire neuron architecture that achieves unprecedented power efficiency, spiking frequency, and energy per spike. Second, we have implemented a tunable threshold mechanism that allows dynamic adjustment of neuron firing behavior while maintaining stability. Third, we provide a comprehensive analysis of the circuit's resilience to process-voltage-temperature (PVT) variations, demonstrating robust performance across a wide operating range. Fourth, we have optimized the design to enable reliable operation with varying input spike pulse widths, supporting diverse application requirements. Our findings pave the way for the integration of novel neuron circuits in next-generation neuromorphic systems, addressing the growing demand for energy-efficient and high-speed computational solutions.

The remainder of this paper is organized as follows: Section II introduces the device structure, operation modes, and models of the S-FED technology. Section III presents the circuit architecture and detailed simulation results. Section IV provides a comprehensive analysis of the circuit's behavior under process, voltage, and temperature variations. Finally, Section V concludes the paper with a summary of findings and future implications for neuromorphic computing systems.

## II. DEVICE STRUCTURE, OPERATION MODES, AND MODELS

The S-FED (Side-Contacted Field-Effect Diode) boasts unique attributes that make it exceptionally well-suited for applications demanding low power consumption and high-speed switching. Key features include a high on-current to off-current ($I_{ON}/I_{OFF}$) ratio, minimized intrinsic gate delay, and reduced parasitic capacitance, all of which contribute to its efficiency and versatility [20-24].

As illustrated in Figure. 1, the S-FED features a dual-gate architecture with two distinct gates—the Gate-Source (GS) and Gate-Drain (GD)—positioned over the channel to provide enhanced control. This gate configuration eliminates the need for separate fabrication of p-MOS, n-MOS and diode components, significantly simplifying circuit implementation. Additionally, the device incorporates heavily doped source and drain regions, each accompanied by reservoir areas beneath them, which improve the $I_{ON}/I_{OFF}$ ratio even in the nanoscale regime. The schematic and dimensional layout of the S-FED are detailed in Figure. 1.

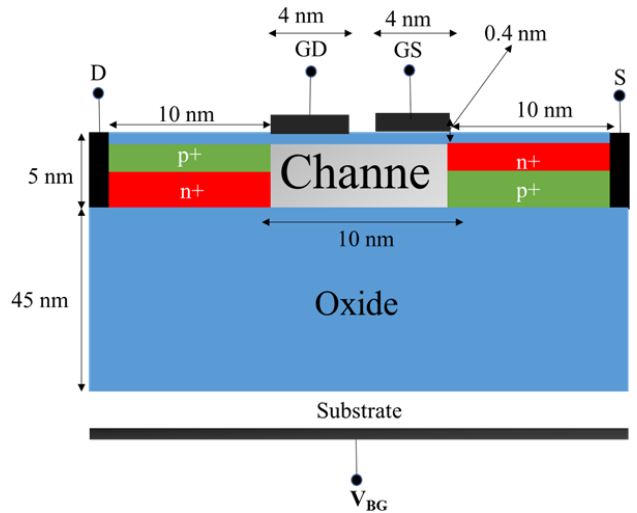

**FIGURE 1.** Structural schematic of the nanoscale S-FED device.

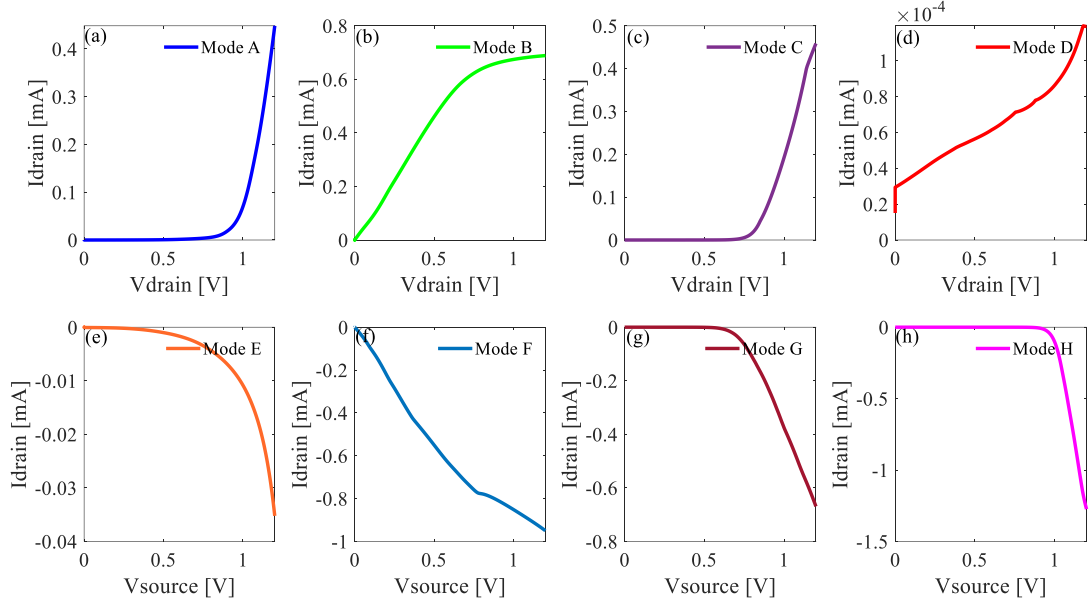

**FIGURE 2.** I-V characteristics of S-FED in different operation modes. (a)-(d) shows mode A-D (VDS>0). (e)-(f) shows mode E-F (VDS<0).

The doping concentration for the source/drain regions and their reservoirs is set to $10^{21} cm^{-3}$, while the channel remains intrinsically doped. The gates have a work function of 4.7 eV, and the power supply voltage ($V_{DD}$) and bulk voltage ($V_{BG}$) are configured at 1.2 V and 0.6 V, respectively. Simulations are conducted at a standard operating temperature of 27°C using the SILVACO TCAD tool to model the integration of the S-FED into an integrate-and-fire (IF) neuron circuit [25].

TABLE 1
S-FED MODES, STATES, AND STRUCTURE IN DIFFERENT BIASES OF GATES AND DRAIN-SOURCE.

| $V_{DS}$ | $V_{GS}$ | $V_{GD}$ | Mode | State | Structure (Drain to Source) |
|---|---|---|---|---|---|
| >0 | >0 | <0 | A | ON | $P^+PNN^+$ |
| >0 | >0 | >0 | B | ON | $P^+NNN^+$ |
| >0 | <0 | <0 | C | ON | $P^+PPN^+$ |
| >0 | <0 | >0 | D | OFF | $P^+IIN^+$ |
| <0 | >0 | <0 | E | ON | $P^+PNN^+$ |
| <0 | >0 | >0 | F | ON | $P^+NNN^+$ |
| <0 | <0 | <0 | G | ON | $P^+PPN^+$ |
| <0 | <0 | >0 | G | ON | $P^+NPN^+$ |

To achieve accurate results, the simulations employ a range of advanced physical models, including Klaassen Band-to-Band Tunneling (BBT.KL), concentration- and field-dependent mobility (Conmob and Fldmob), Shockley-Read-Hall (SRH) and Auger recombination, bandgap narrowing, and the Lombardi CVT model. The Newton numerical method is used to solve the system equations, offering significant advantages in convergence. This method allows small incremental voltage changes to provide better initial approximations at each step, minimizing convergence issues and enhancing simulation accuracy.

Table 1 provides an overview of the S-FED's performance across different operational regimes. As shown in both Table 1 and Figure. 2, the S-FED demonstrates switchable behavior between the ON and OFF states when the drain-source voltage $V_{DS}$ is positive. However, when $V_{DS}$ is negative, the device remains perpetually in the ON state. This constant ON-state behavior arises from tunneling effects and the short-channel effect, rendering the S-FED unsuitable for switching applications under these conditions

### III. CIRCUIT SECTION: ARCHITECTURE AND SIMULATIONS

In this section, we investigate the behavior of the integrate-and-fire (IF) neuron circuit utilizing nanoscale S-FEDs, as shown in Figure 3. After this analysis, we present mixed-mode simulation results for the circuit. We then demonstrate the threshold tunability of the IF neuron circuit and explore its operation under different pulse width variations of the input signal. Finally, we compare the performance of the nanoscale S-FED-based IF neuron model with the latest state-of-the-art neuron models.

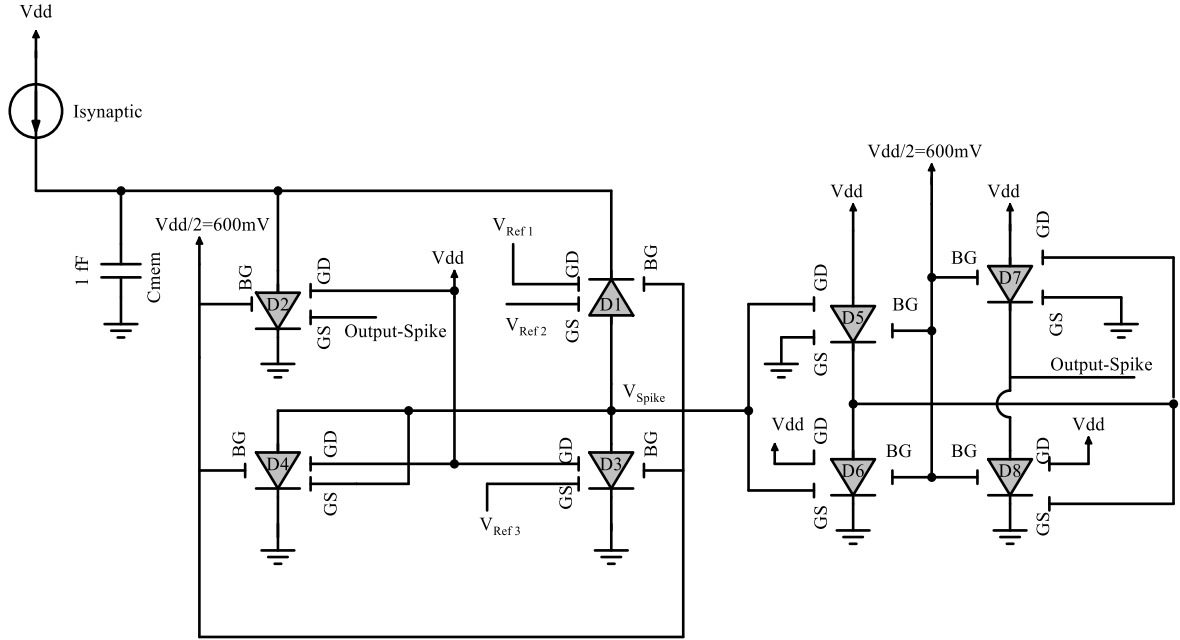

**FIGURE 3.** Schematic of proposed IF neuron circuit using S-FED.

### A. CIRCUIT DESIGN AND SIMULATION

Figure 3 illustrates the mechanism by which synaptic spikes ($I_{synaptic}$) initiate the charging of the membrane capacitor ($C_{mem}$). When the membrane potential exceeds the threshold of the first S-FED (D1), it activates the device, allowing a substantial current to flow through this pathway. This current is subsequently converted into a voltage by the third S-FED (D3). Meanwhile, the fourth S-FED (D4) operates in mode B, mimicking a diode-connected n-MOS transistor to reset the spike node with a consistent current. To stabilize and buffer the spike node, two back-to-back inverters (D5, D6, D7 and D8) are employed. These inverters suppress any potential below 500 mV to zero while amplifying potentials above 500 mV to 1.2 V. This buffering mechanism is crucial for eliminating unwanted spikes and enhancing the amplitude of valid spikes, ensuring reliable triggering of the second S-FED (D2). The Output-Spike node activates D2, which is also biased in mode B, facilitating the discharge of the membrane capacitor. This process resets the neuron circuit, preparing it to receive the next synaptic spike. The voltages applied to the gate-drain and gate-source ($V_{GD}$ and $V_{GS}$) of the first S-FED (D1) are set to approximately $V_{Ref1}$ = 800 mV and $V_{Ref2}$ = 100 mV, respectively. These references bias D1 into mode H, in which it functions akin to a diode (Figure. 2(h)). Additionally, a 400 mV is applied to $V_{Ref3}$ to facilitate D3's behavior as a resistor for current-to-voltage conversion. The entire IF neuron circuit operates at a supply voltage ($V_{dd}$) of 1.2V, with the back-gate voltage ($V_{BG}$) set to $V_{dd}/2$. Notably, the amplitude of synaptic input spikes ($I_{synaptic}$) is maintained at 250 nA, and the membrane capacitor has a capacitance of 1fF. Simulation results of the proposed IF neuron circuit are presented in Figure. 4. Figure 4 (b) illustrates the generation of an action potential following five input stimulations, while Figure. 4 (c) demonstrates the charging and discharging behavior of the membrane node. Notably, upon receiving the fifth synaptic input spike, the neuron fires an action potential at the Output-Spike node, showcasing the circuit's effective operation.

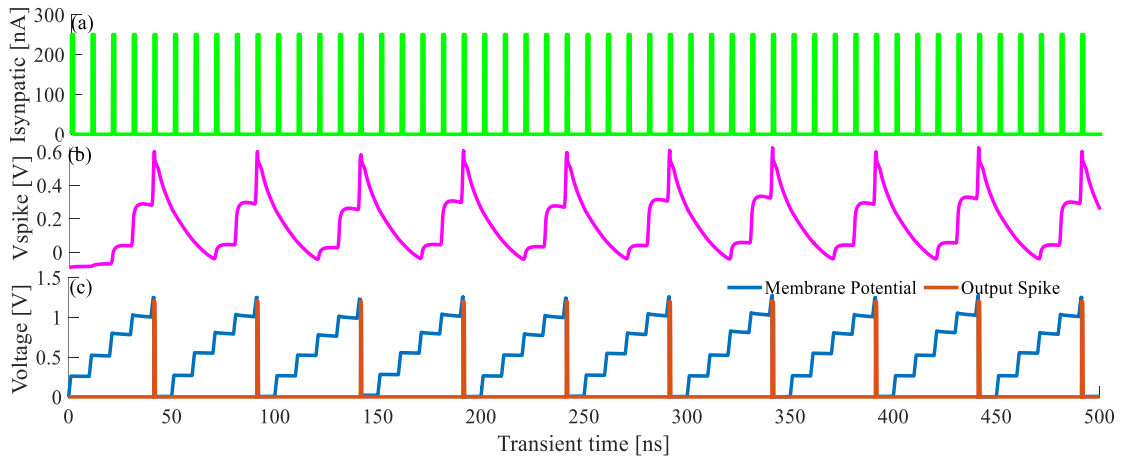

**FIGURE 4.** Simulation results of proposed IF neuron circuit. (a) Illustrates the synaptic current (input spikes). (b) Depicts the voltage of the spike node over time. (c) Shows the charging behavior of the membrane capacitor and the occurrence of output spikes.

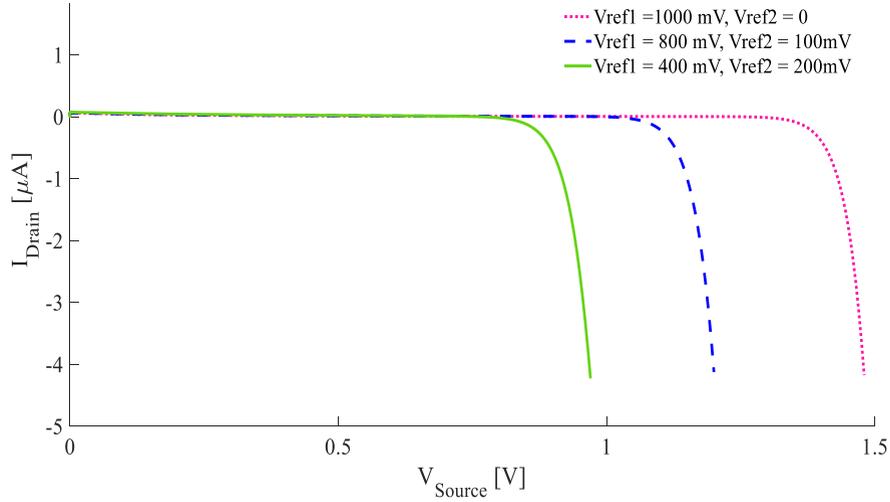

**FIGURE 5.** Threshold tunability of the side-contacted Field Effect Diode (S-FED) with variations in gates bias ($V_{GD}$ and $V_{GS}$). The S-FED in this simulation is operated in mode H.

### B. THRESHOLD TUNABILITY AND CIRCUIT BEHAVIOR UNDER VARITION OF PULSE WIDTH OF INPUT SIGNAL

One of the key features of our IF neuron architecture is the tunability of its threshold. This is achieved by applying appropriate bias voltages to the gates of D1. These bias voltages modulate the potential barrier in the channel of the S-FED, thereby altering its threshold. As shown in Figure. 5 and Figure 6 (a), when $V_{ref1}$ and $V_{ref2}$ are set to 400 mV and 200 mV, respectively, the threshold of D1 is 800 mV. Increasing $V_{ref1}$ to 800 mV and decreasing $V_{ref2}$ to 100 mV raises the potential barrier in the channel, shifting the threshold of D1 to 1.1 V (Figure 5 and Figure 6 (b)). Furthermore, increasing $V_{ref1}$ to 1000 mV and reducing $V_{ref2}$ to 0 V further increases the potential barrier, resulting in a threshold shift to 1.4 V (Figure 5 and Figure 7 (b)).

Another key characteristic of neuron circuits is its output response to varying input pulse widths. Figure. 7 illustrates the output spikes of our neuron architecture in response to different input pulse widths.

In Figures 7(a) and 7(e), when the input signal's pulse width is set to 0.5 ns, the membrane potential reaches the threshold of the IF neuron after 10 input spikes. As the pulse width increases (Figure 7(b)), the membrane capacitor charges more quickly, causing the membrane potential to reach its threshold sooner. Consequently, the IF neuron fires after only five input spikes, as shown in Figure 7(f). This trend continues with further increases in pulse width, as demonstrated in Figures 7(c) and 7(d), along with their corresponding responses in Figures 7(g) and 7(h).

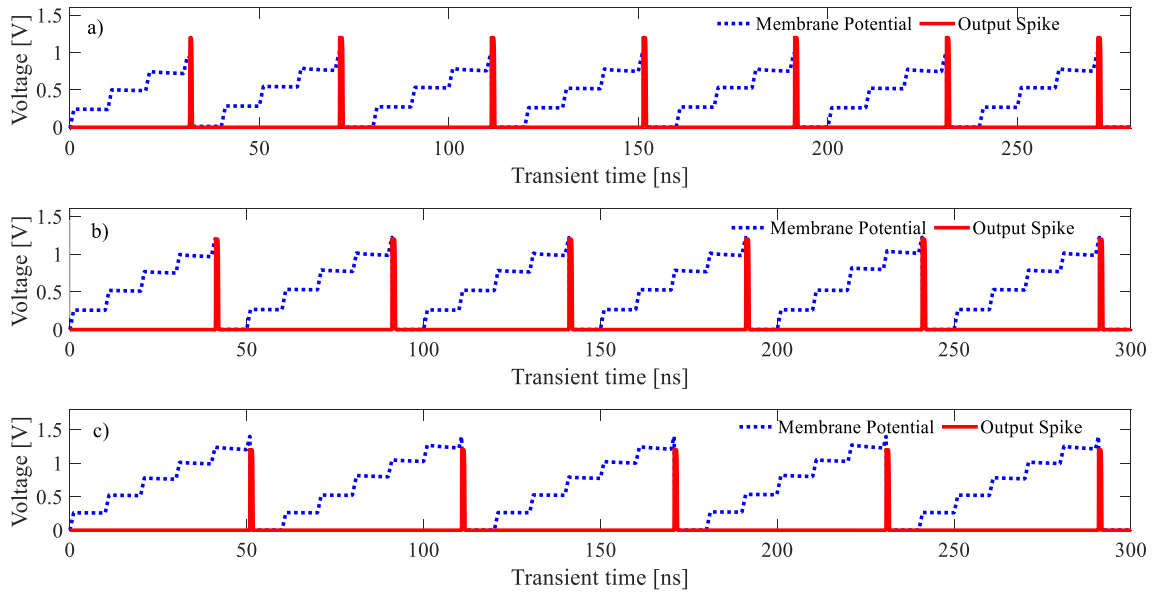

**FIGURE 6.** Membrane potential and output spike voltage as a function of time for different values of $V_{Ref1}$ and $V_{Ref2}$. (a) Transient simulation of the IF neuron circuit with Vref1=400 mV and Vref2=200 mV. (b) Transient simulation of the IF neuron circuit with Vref1=800 mV and Vref2=100 mV. (c) Transient simulation of the IF neuron circuit with Vref1=1000 mV and Vref2=0.

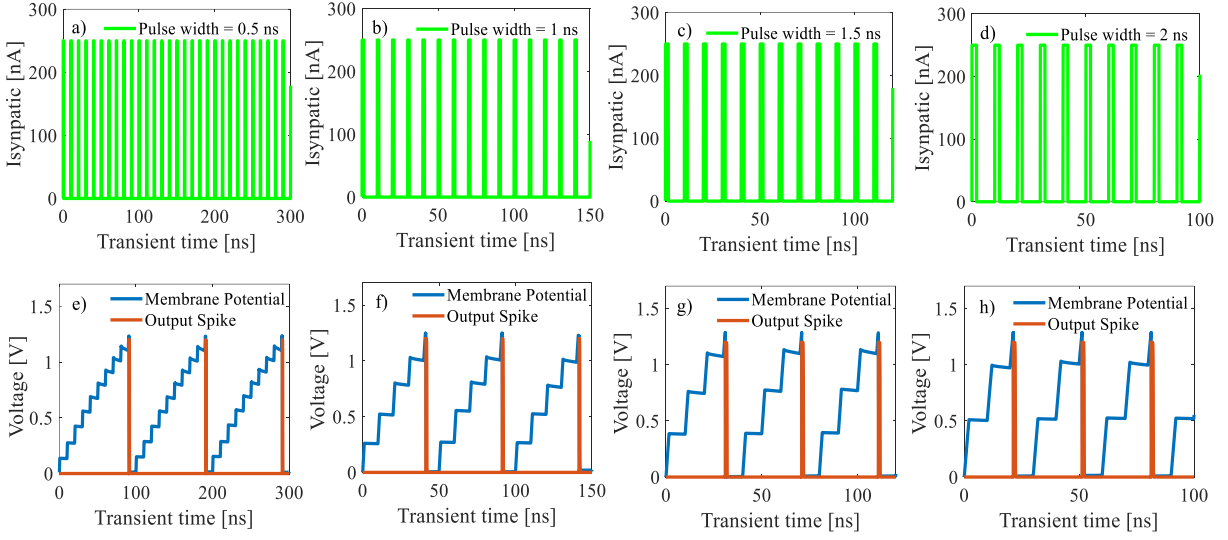

**FIGURE 7.** (a-d) Synaptic input current as a function of time for varying pulse widths. (e-h) Transient simulation results showing the membrane potential and output spike node of the IF neuron circuit, where (e-h) correspond to the simulation results for the synaptic input currents shown in (a-d), respectively. varying pulse width from 0.5 ns to 2 ns steps

### C. PERFORMANCE ANALYSIS OF THE PROPOSED IF NEURON CIRCUIT

Using Eq. (1), the energy consumption per spike for the proposed neuron circuit was calculated to be approximately 0.96425 fJ, a value significantly lower than that of existing state-of-the-art neuron circuits. Additionally, the spiking frequency of the proposed circuit reaches approximately 20 MHz, surpassing that of its counterparts. DC simulations were performed, and the DC power was computed using Eq. (2), yielding a lower value compared to other recent advanced neuron circuits. Table 2 highlights the superior performance of the proposed IF neuron circuit across key metrics compared to other state-of-the-art designs. Its remarkable energy efficiency makes it highly suitable for energy-sensitive applications, while the high spiking frequency supports rapid signal processing, essential for complex computations and real-time artificial neural network operations. The circuit's low DC power consumption underscores not only its efficiency but also its viability for dense neural computing systems where power management is critical. These attributes collectively establish the proposed IF neuron circuit as a significant advancement in neuromorphic engineering.

$$E_S[\text{J/spike}] = \frac{1}{Tf_s}\int_0^T I_{synaptic} V_{membrane} dt \quad \text{Eq. (1)}$$

$$Power\ consumption = V_{dd}I_{dd} = V_{dd}(I_{D5} + I_{D7}) \quad \text{Eq. (2)}$$

TABLE 2
COMPARISON OF NEURON CIRCUIT PERFORMANCE. THREE SETS OF EXPERIMENTAL DATA WERE REPORTED FOR DEVICE PERFORMANCE, WHILE CIRCUIT-LEVEL PERFORMANCE WAS OBTAINED THROUGH SIMULATION SOFTWARE. ALL RESULTS WERE MEASURED AT ROOM TEMPERATURE

| Year | Ref. | Experimental / Simulation | Neuron model | Device type | Technology [nm] | Power [μW] | Energy [J/spike] | Spike Frequency |
|---|---|---|---|---|---|---|---|---|
| 2016 | [13] | Simulation | IF | Phase Change+ CMOS | 90 | - | 3×10⁻¹¹ | 10 Hz – 10 MHz |
| 2017 | [14] | Experimental | LIF | Floating-Body MOSFET | 32 | - | 3.5×10⁻¹¹ | 20 MHz |
| 2019 | [16] | Simulation | LIF | FinFET | 100 | - | 6.3×10⁻¹⁵ | 1.9 MHz |
| 2020 | [17] | Experimental | LIF | BTBT | 32 | - | 3.2×10⁻¹⁵ | 150 kHz |
| 2020 | [26] | Simulation | IF | FBFET | 50 | 7 | 2.9×10⁻¹⁵ | 20 kHz |
| 2023 | [18] | Simulation | IF | RRAM | 65 | - | 1.5×10⁻¹⁵ | 277 kHz -3 MHz |
| 2024 | [19] | Simulation | LIF | TG-V-FBFET | 40 | - | 0.13×10⁻¹² | - |
| **2025** | **This work** | **Simulation** | **IF** | **S-FED** | **10** | **0.044** | **0.96425 ×10-15** | **20 MHz** |

## D. METHODS

To simulate this circuit, we first conduct device- and circuit-level simulations using technology computer-aided design (TCAD) tools, specifically employing a semiconductor drift-diffusion solver to analyze the characteristics of the IF neuron circuit based on the DS-S-FED. The process begins with the calibration of an analogous DS-S-FED, where the geometrical parameters are selected to align with the high-performance (HP) logic technology outlined in the ITRS roadmap60. The physical model used for simulating this circuit is identical to the one described for DS-S-FED in Section II, Part A [28-29].

Mixed-mode circuit simulators handle data transfer between device and circuit simulation in a structured manner. Initially, a physics-based device simulator is employed to compute the electrical characteristics of individual components. These characteristics are then passed to a device modeling and parameter extraction tool. This tool generates compact models, and simplified representations of the device's behavior, which are subsequently used by the circuit simulator. During both steady-state and transient analysis, Newton's algorithm is used to solve for circuit voltages.

To ensure stability and accuracy, several parameters are defined. The maximum voltage change allowed between iterations is limited to 0.1V, and specific accuracy targets are set for voltage calculations: a relative accuracy of 0.1 for steady-state analysis and a much stricter 0.001 for transient analysis. For transient simulations, which analyze behavior over time, a minimum time step of 50 attoseconds is used, and the local truncation error is limited to 0.001. A maximum of 40 iterations are permitted for both steady-state and transient analysis. This particular setup is being used to simulate a neuron circuit driven by a fast current pulse. The input current, ranging from 0 to 250 nA, is characterized by a sharp 1 picosecond rise and fall time. A very fine time interval of 0.01 picoseconds is used to capture subtle changes in the output voltage signal. This highlights the capability of the simulator to handle both detailed device physics and fast transient behavior in a circuit context.

## IV. PROCESS-VOLTAGE-TEMPERATURE (PVT) ANALYSIS

It excels in integrating and firing neuron circuits to operate reliably under challenging physical and environmental conditions, a process referred to as process-voltage-temperature (PVT) analysis. This section explores the impact of variations in channel length, supply voltage, and temperature on the performance of the nanoscale S-FED-based neuron circuit. Among these factors, channel length scaling plays a crucial role in maintaining stability. Specifically, designing circuits that function reliably despite undesirable variations in channel length during the fabrication process is particularly significant. As illustrated in Figure. 8(a), an increase in channel length raises the threshold of the integrate-and-fire (IF) neuron circuit, as carriers require higher energy to transmit through the longer channel. Figure. 8(b) highlights that altering the channel length from 7.5 nm to 15 nm results in only a minor change in spike amplitude, approximately 7 mV, which ensures the stability of spike amplitude in the IF neuron architecture. Figure 8(c) demonstrates that variations in channel length have no impact on the spiking frequency of the IF neuron circuit. Finally, Figure. 1(d) illustrates the impact of channel length variation on the static power consumption of the IF neuron circuit. As shown, reducing the channel length from 15 nm to 7.5 nm increases static power consumption from 14 nW to 60 nW. This rise is attributed to the increased current through the inverters as the channel length decreases.

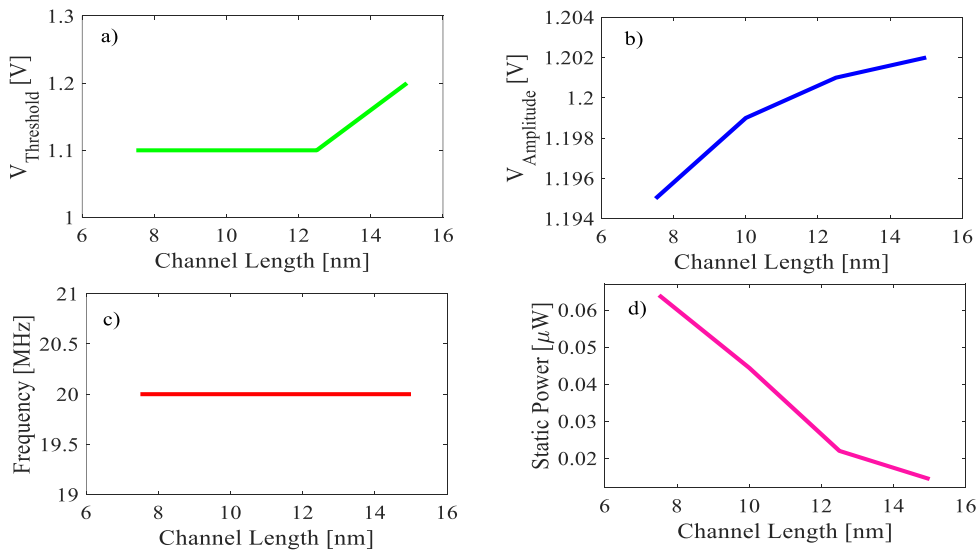

**FIGURE 8.** effect of channel length variation on the performance of nano-scale S-FED based neuron circuit. (a) Impact of channel length variations on the threshold voltage of S-FED based IF neuron circuit. (b) Impact of channel length variations on the amplitude of spike of S-FED based IF neuron circuit. (c) Impact of channel length variations on the spiking frequency of S-FED based IF neuron circuit. (d) Impact of channel length variations on the static power of S-FED based IF neuron circuit.

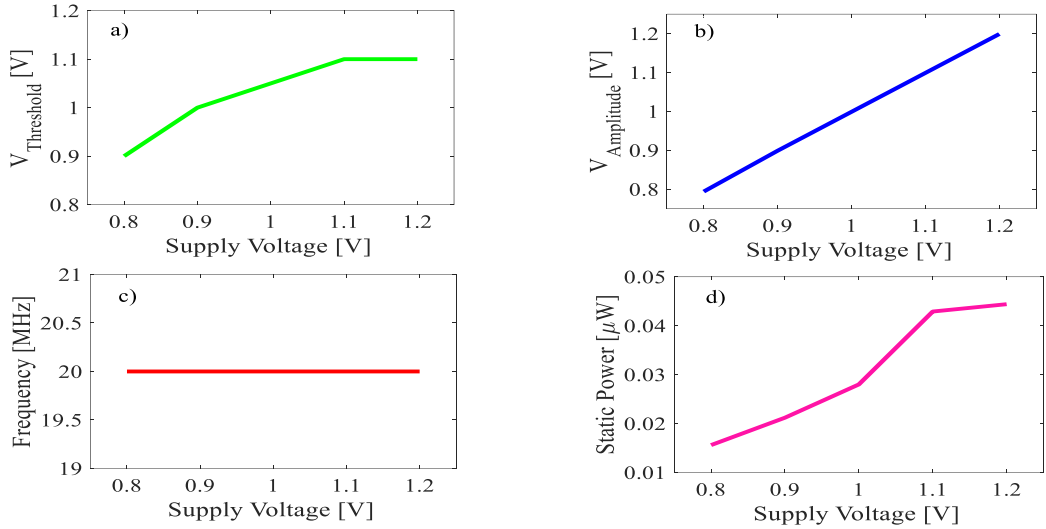

**FIGURE 9.** Effect of supply voltage variation on the performance of nanoscale S-FED based neuron circuit. (a) Effect of supply voltage changes on the threshold voltage of S-FED based neuron circuit. (b) Effect of supply voltage changes on the spikes' amplitude of S-FED based neuron circuit. (c) Effect of supply voltage changes on the spiking frequency of S-FED based neuron circuit. (d) Effect of supply voltage changes on the static power of S-FED based neuron circuit.

Another critical factor affecting the performance of the IF neuron circuit is changes in supply voltage. Maintaining an exact supply voltage during operation can be challenging, as noise or other disturbances may alter its value. As shown in Figure. 9(a), variations in supply voltage from 0.8 V to 1.2 V shift the threshold of the IF neuron circuit by approximately 200 mV. Figure. 9(b) demonstrates that an increase in supply voltage is directly related to an increase in spike amplitude, as the supply voltage defines the maximum voltage the spikes can reach. However, this change does not impact the spiking frequency of the IF neuron architecture, as depicted in Figure. 9(c). Finally, Figure. 9(d) illustrates the relationship between supply voltage and static power. As observed, higher supply voltages result in increased static power consumption due to the direct relationship between the static power of the inverters used in this architecture and the supply voltage (Eq. (2)).

Another crucial factor that can influence the performance of the IF neuron circuit is temperature. The circuit's ability to maintain stable integration and firing operations across a wide temperature range makes it well-suited for use in diverse environments. To simulate the effect of temperature on circuit behavior, Arora model effects are incorporated into the simulation models for analyzing of influence of temperature changes on the performance of the S-FED based IF neuron circuit. As shown in Figure. 9(a-d), temperature variations from -40°C to 120°C have no significant impact on the performance of the IF neuron architecture, making it suitable for applications in harsh environments. This remarkable stability of the performance of S-FED-based IF neuron architecture is related to the high carrier concentration of source and drain region and the robust feedback mechanism in the spike generation part of the circuit. These features help maintain reliable neural behavior across varying environmental conditions, making the circuit suitable for real-world neuromorphic applications.

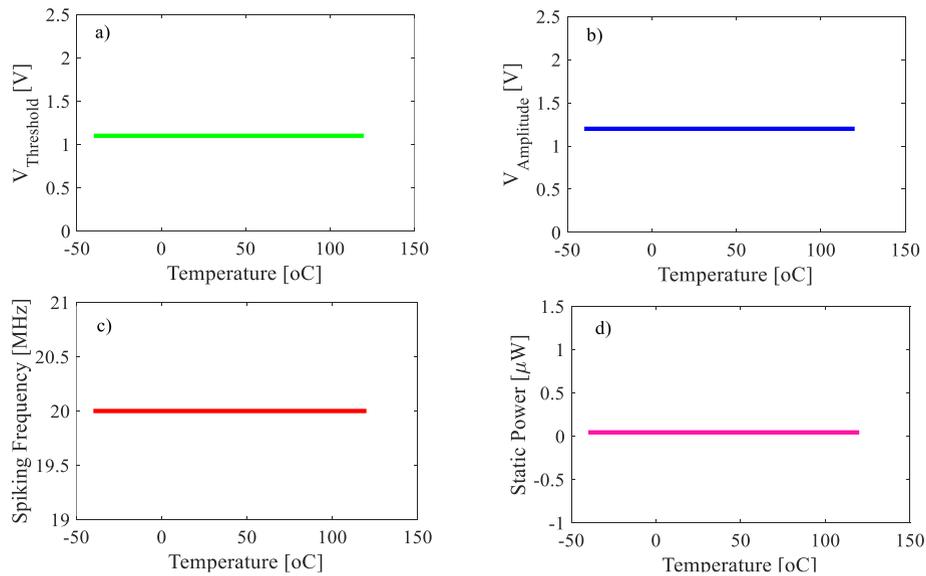

**FIGURE 10.** Effect of temperature variation on the performance of nanoscale S-FED based neuron circuit. (a) Effect of temperature variation on the threshold voltage of S-FED based neuron circuit. (b) Effect of temperature variation on the spikes' amplitude of S-FED based neuron circuit. (c) Effect of temperature variation on the spiking frequency of S-FED based neuron circuit. (d) Effect of temperature variation on the static power of S-FED based neuron circuit.

## V. CONCLUSION

This paper explores the characteristics of S-FED, demonstrating its functionality across various modes and emphasizing its suitability for designing IF neuron circuits. We then introduce our proposed neuron architecture and explain how the circuit implements the IF neuron model. Additionally, we present simulation results showcasing the adjustable threshold of the nanoscale S-FED-based integration-and-firing neuron circuit. The operation of the neuron circuit under varying pulse widths of input signals is also illustrated. In the subsequent section, we compare the performance of the nanoscale S-FED-based IF neuron with other state-of-the-art neuron architectures. The results indicate that our design outperforms its counterparts in terms of static power consumption, energy per spike, and spiking frequency. Finally, we examine the effects of channel length variations, supply voltage fluctuations, and temperature changes on the performance of the S-FED-based IF neuron model. The findings reveal that these parameters have minimal impact on the circuit's overall functionality.